\documentclass[aps,a4paper,byrevtex,twocolumn,showkeys,longbibliography,floatfix]{revtex4-1}
\usepackage[T1]{fontenc}
\usepackage{amsmath}
\usepackage{amssymb}
\usepackage{graphicx}
\usepackage{xcolor}
\usepackage[colorlinks=true,hyperfootnotes=true,breaklinks=true,backref=none]{hyperref}

\begin{document}
\title{Paradox of integration---Cellular automata approach}

\author{K.~Malarz}
\thanks{\includegraphics[scale=0.6]{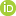}~\href{https://orcid.org/0000-0001-9980-0363}{0000-0001-9980-0363}}
\email{malarz@agh.edu.pl}
\affiliation{\href{http://www.agh.edu.pl/}{AGH University of Science and Technology},
\href{http://www.pacs.agh.edu.pl/}{Faculty of Physics and Applied Computer Science},\\
al. Mickiewicza 30, 30-059 Krak\'ow, Poland.}

\author{K.~Ku{\l}akowski}
\thanks{\includegraphics[scale=0.6]{orcid_16x16.png}~\href{https://orcid.org/0000-0003-1168-7883}{0000-0003-1168-7883}}
\email{kulakowski@fis.agh.edu.pl}
\affiliation{\href{http://www.agh.edu.pl/}{AGH University of Science and Technology},
\href{http://www.pacs.agh.edu.pl/}{Faculty of Physics and Applied Computer Science},\\
al. Mickiewicza 30, 30-059 Krak\'ow, Poland.}


\date{\today}

\begin{abstract}
We discuss the self-deprecating strategy introduced by Peter Blau as one of stages of the process of social integration. Recently we have introduced a two-dimensional space of status, real and surface one ($A$ and $B$), and we have demonstrated that with this setup, the self-deprecating strategy efficiently prevents the rejection \citetext{K.~Malarz and K.~Ku{\l}akowski, International Journal of Modern Physics C {\bf 30}, 1950040 (2019)}. There, the process of reducing the conflict was described by master equations, i.e. a set of differential equations describing evolution of density $v(A,B)$ of actors of status $(A,B)$. Here we reformulate the problem in terms of probabilistic cellular automata. The obtained results for number $n(A,B)$ of actors of status $(A,B)$ are qualitatively the same as in the previous approach, both for synchronous and asynchronous version of the automaton. Namely, an enhancement of the surface status compensates a deficiency of the real one. 
\end{abstract}

\maketitle


\section{Introduction}

According to Peter Blau, social integration of a group of adults contains two stages \cite{Blau}. In first stage, actors demonstrate their strongest points, to achieve social status as high as possible. This stage can be painful for some, who have no virtues to present; they respond with fear and hostility.  As Blau puts it: ``The more successful A is in impressing B and earning B's high regard, the more displeasure he causes to C, whose relative standing in the eyes of B has suffered. All group members simultaneously play the role of A, B, and C in this schema, which greatly complicates the competitive process'' \cite[p. 44]{Blau}. This leads to a paradoxical phenomenon: persons most skilful, intelligent and physically attractive meet with rejection and hostility. To neutralize this, in the second stage intelligent persons know and apply a clever strategy: they seemingly reduce their advantage, demonstrating their weak points in less important aspects of status. To cite Blau again: ``Having first impressed us with his Harvard accent and Beacon Hill friends, he may later tell a story that reveals his immigrant background'' \cite[p. 48]{Blau}. It is worth to mention that this strategy is not efficient if applied by an actor too weak or too strong. In first case, it is read as a fake suggestion of non-existent strong points \cite[pp. 48-49]{Blau}, in the second---as an arrogant demonstration of lack of understanding of difficulties of the others' life \cite{rmn}. As a fairly complex social process, the social integration is a promising playground for an interdisciplinary research, between social sciences and computational modelling.

The self-deprecating strategy (SDS) depicted by Blau has been the subject of a series of recent papers \cite{m1,m2,1903.04291}. Yet, it is only in the last \cite{1903.04291} where the model results successfully reproduced the efficiency of SDS as a tool to reduce the fear-driven rejection. This  was achieved by an introduction of a two-dimensional space of status, with real and surface axes. Along these two axes, two processes were competing: fear-induced rejection equivalent to a shift down along the real axis, and preventive praising, which drives the status of an opponent up along the surface axis. The probabilities of these two strategies ($\alpha$ and $1-\alpha$, respectively) have been used as parameters; below we keep the same notation. The formalism applied was a set of differential equations, with both probability distribution of agents in the space of status and the related cumulative distribution involved as variables. In this sense, the description presented in Ref.~\onlinecite{1903.04291} was non-local; the time evolution of the status of actors depended not only on their direct neighbours, but also on those fairly distant on the status plane.

To which extent social interactions are active between individuals of clearly different social status depends on the context. Excerpts from the Blau book \cite{Blau} given above indicate a living room or another informal gathering where new acquaintances are made. Even there, the answer relies on local culture \cite[p.  1401]{weber}, \cite{Goody-1971,Ridgeway-Smith-Lovin-1999}. On the other hand, both in Ref.~\onlinecite{1903.04291} and here we have in mind SDS applied in a scale of groups and not only individuals. One of examples quoted in Ref.~\onlinecite{1903.04291} is the glorification of working class in communist countries, attributing the role of dictators to proletariat \cite{manikom};  the scale of this manipulation was transnational.  Yet, it is obvious that in this and similar cases SDS is expected to be more efficient when performed by groups of status at least nominally close to the one of the target group. In particular, the Soviet Politburo composed of aristocrats would be much less credible for working classes.

Here we are interested on the efficiency of SDS if applied only to nearest neighbours in the space of status.  According to the note in the first paragraph, if the difference of statuses of two individual is large the mechanism is less effective. Our aim is to check, how the results of the simulation depend on the assumption on the local character of interactions. To achieve this purpose, the problem is reformulated in the frames of probabilistic cellular automata, the formalism local by definition \cite{Adamatzky-2018,*Wolfram-2002,*Ilachinski-2001,*Hegselmann-2000,*Hegselmann-1998}. Here, the time evolution of the positions of actors in the status plane depends only on their direct neighbours in this plane. Besides this, we keep the time-dependent distribution of actors in the status space as the variable, as was done in Ref.~\onlinecite{1903.04291}. 

\begin{table}
\caption{\label{tab:evolution}Videos showing the status--temporal evolution of the number $n(A,B)$ of actors in status $(A,B)$ for synchronous and asynchronous sites update and various values of $\alpha$, $L_A=40$, $L_B=60$. The final system states are presented in Figs.~\ref{fig:synchronous} and~\ref{fig:asynchronous}.}
	\begin{ruledtabular}
		\begin{tabular}{lr}
		$\alpha$ & URL                                      \\ \hline
		\multicolumn{2}{c}{synchronous update}              \\ \hline
                0.00 & \url{http://www.zis.agh.edu.pl/files/synchro000.gif}\\
                0.25 & \url{http://www.zis.agh.edu.pl/files/synchro025.gif}\\
                0.50 & \url{http://www.zis.agh.edu.pl/files/synchro050.gif}\\
                0.75 & \url{http://www.zis.agh.edu.pl/files/synchro075.gif}\\
                1.00 & \url{http://www.zis.agh.edu.pl/files/synchro100.gif}\\ \hline
		\multicolumn{2}{c}{asynchronous update}             \\ \hline
		0.00 & \url{http://www.zis.agh.edu.pl/files/asynchro000.gif}\\
		0.25 & \url{http://www.zis.agh.edu.pl/files/asynchro025.gif}\\
		0.50 & \url{http://www.zis.agh.edu.pl/files/asynchro050.gif}\\
		0.75 & \url{http://www.zis.agh.edu.pl/files/asynchro075.gif}\\
		1.00 & \url{http://www.zis.agh.edu.pl/files/asynchro100.gif}\\
		\end{tabular}
	\end{ruledtabular}
\end{table}

The automaton rule is checked both for synchronous (parallel) and asynchronous (sequential) version. This is done because we can expect that for intermediate values of the coefficient $\alpha$ the results of the simulations within synchronous and asynchronous scheme are different. This expectation is in agreement with literature. Since publication of \citet{Blok-1999} paper, we are aware that the updating scheme influence the results of simulation based on cellular automata technique. For famous Conway's ``Game of life'' \cite{Berlekamp-1982} automaton changing synchronous to asynchronous updating scheme results in final picture of the lattice similar to the maze instead of well-known lattice of  structures with density of life not exceeding 3\% \cite{Banioli-1991,Antoniuk1998}. Also for Ising model the application parallel or sequential spins updating leads to different results \cite{Skorupa-2012}.  As \citet{Skorupa-2012} put it: ``the problem of updating methods is widely discussed in a recent work on cellular automata, Boolean networks, neural networks, and the so-called agent-based modelling in ecology and sociology \cite{Bandini-2012,Caron-Lormier-2008,Castellano-2009}. It has been shown that the updating scheme can have an enormous influence on the model output \cite{Cornforth-2005}.''   The list of examples of papers devoted to difference between synchronous and asynchronous scheme of sites updating may be extended further, for instance for paper published in the journal devoted to biosystems \cite{Schonfisch-1999}. Also in Ref.~\onlinecite{Skorupa-2012} authors speculate, that differences in system behaviours in sequential or synchronous updating may depend on existence (or not) of the equilibrium, i.e. on satisfying (or not) the detailed balance condition by the updating rules. As automaton rules are asymmetric---the flow of actors may be from bottom to top (Eq.~\eqref{eq:sds}) and from right to left (Eq.~\eqref{eq:nosds})---the detailed balance conditions are violated and thus the order of site updates may influence the results. These arguments incline us to check how the results of our simulations depend on the details of updating the cell states.

In the next section, the automaton is described in detail. Third section is devoted to our numerical results, presented in form of computer animations. The same way of presentation was used in Ref.~\onlinecite{1903.04291}. A short discussion is closing the text.

\section{Model}

\begin{figure*}
\includegraphics[width=0.30\textwidth]{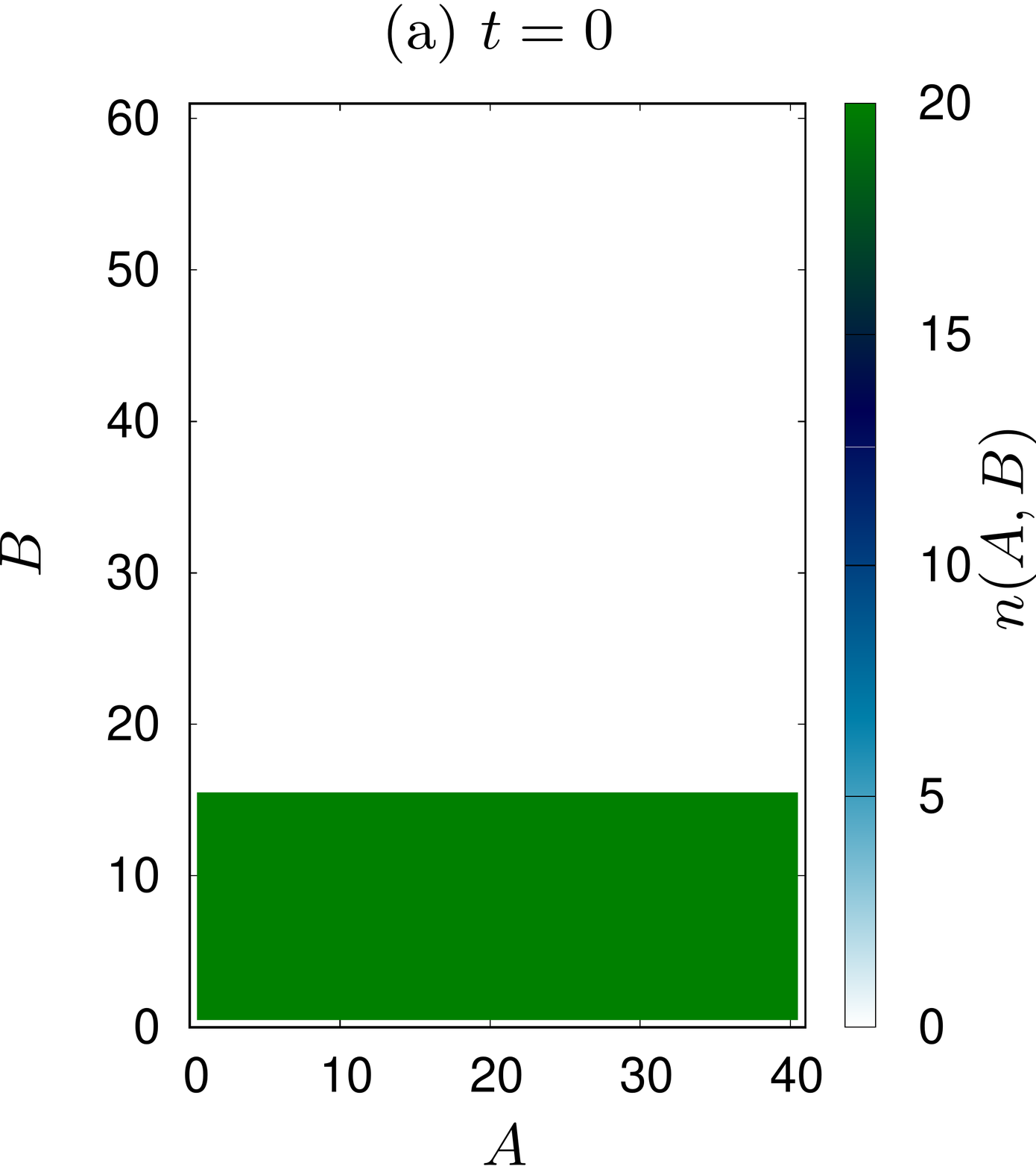}\hfill\includegraphics[width=0.30\textwidth]{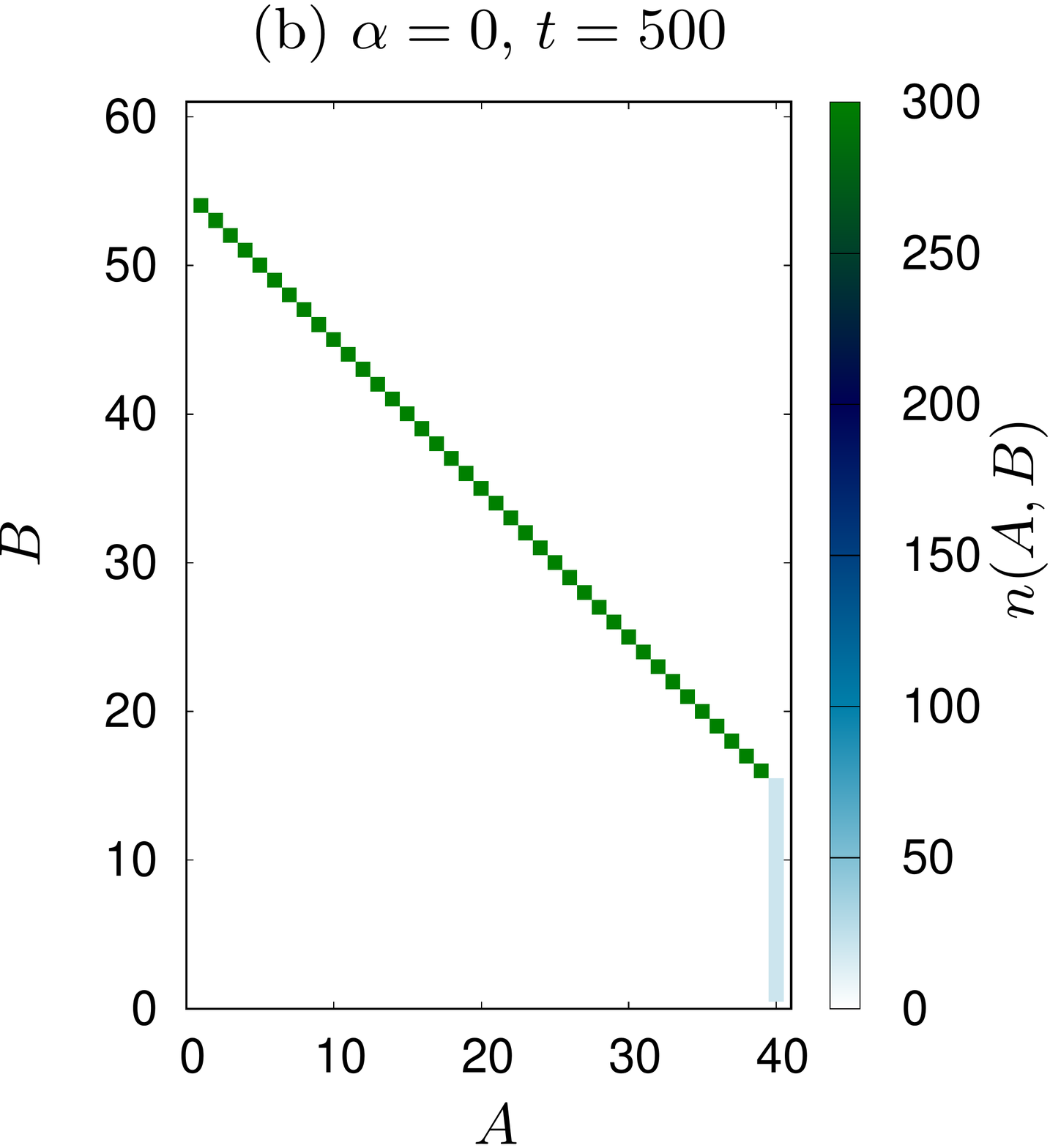}\hfill\includegraphics[width=0.30\textwidth]{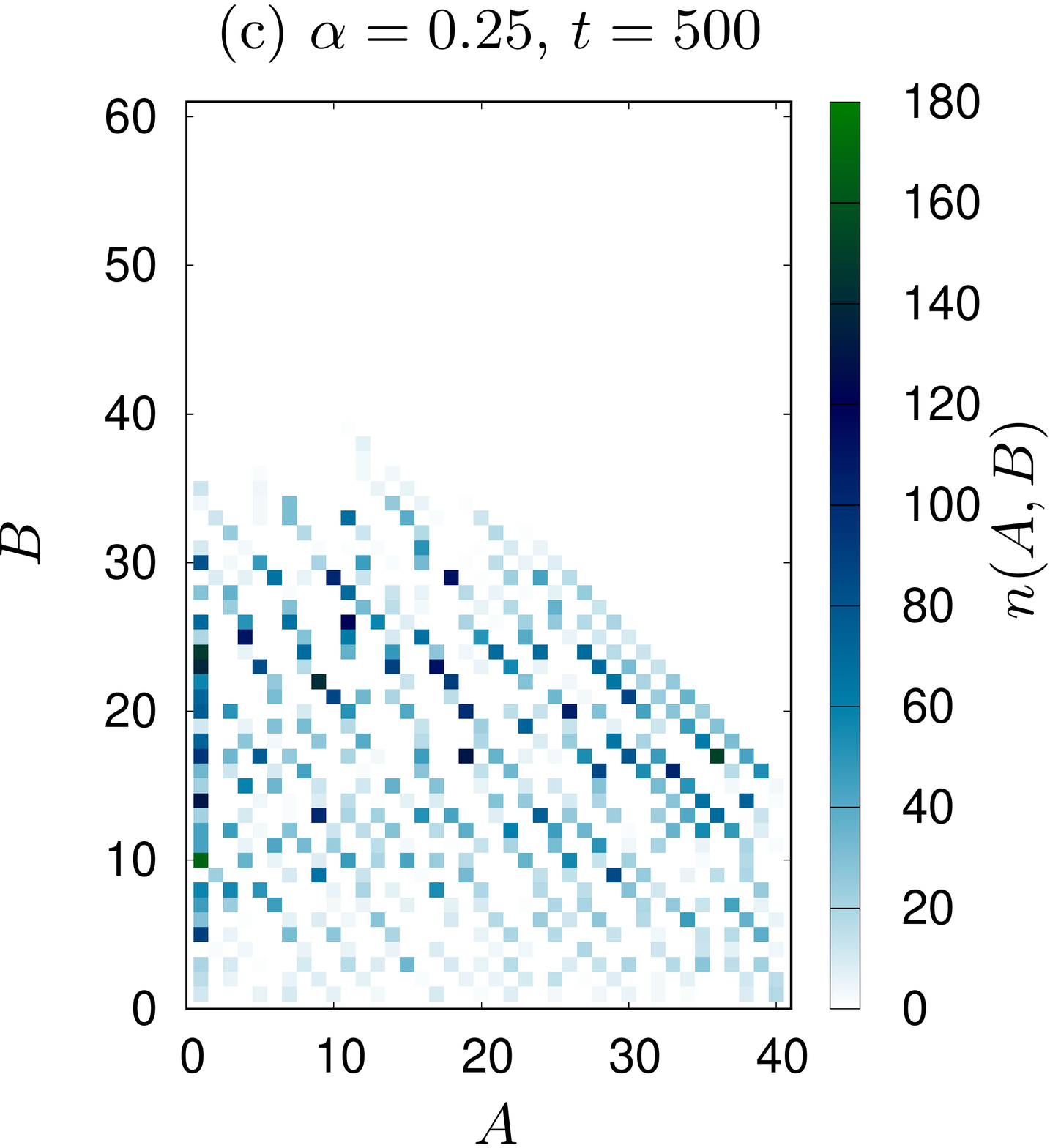}\\
\includegraphics[width=0.30\textwidth]{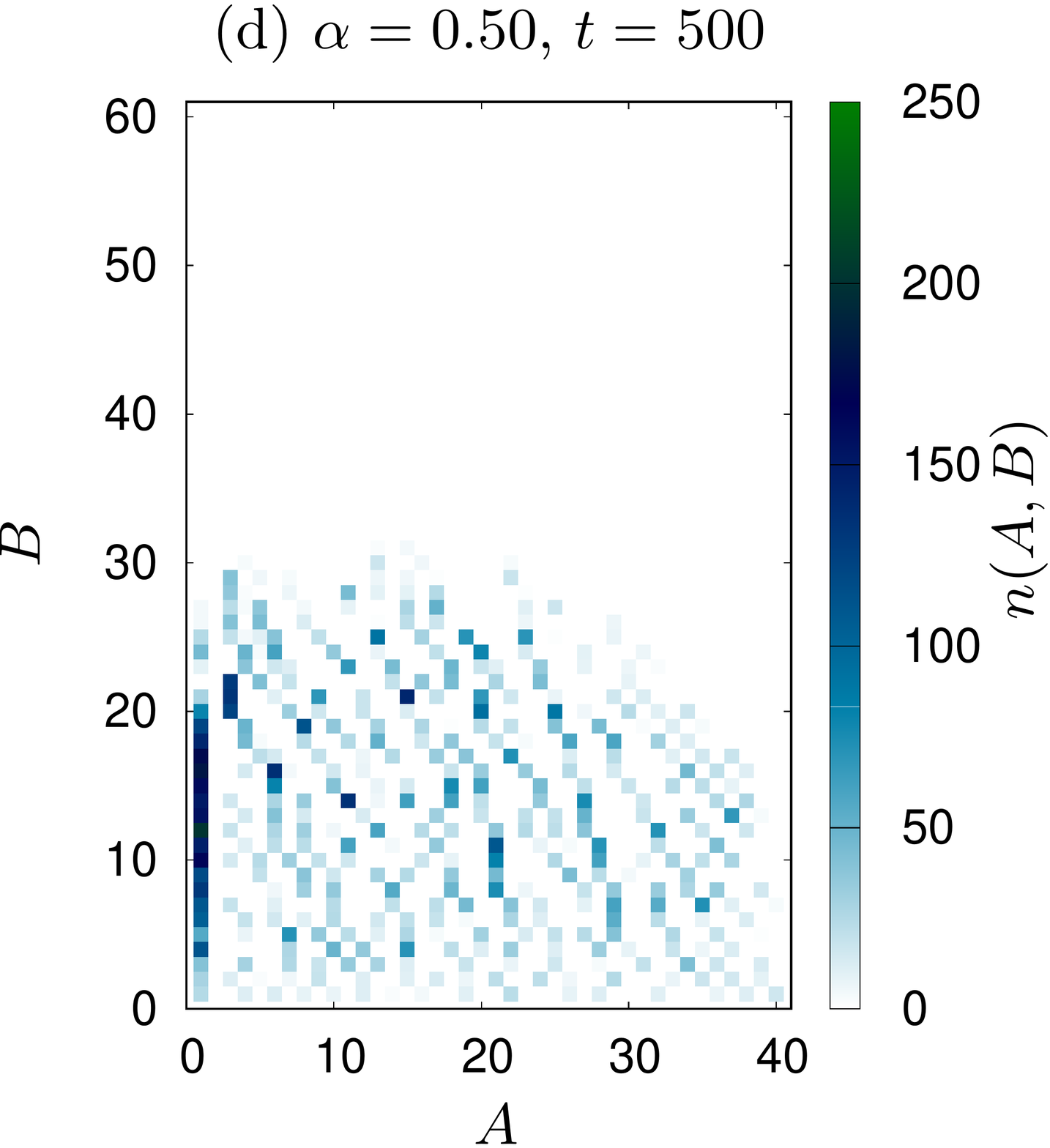}\hfill\includegraphics[width=0.30\textwidth]{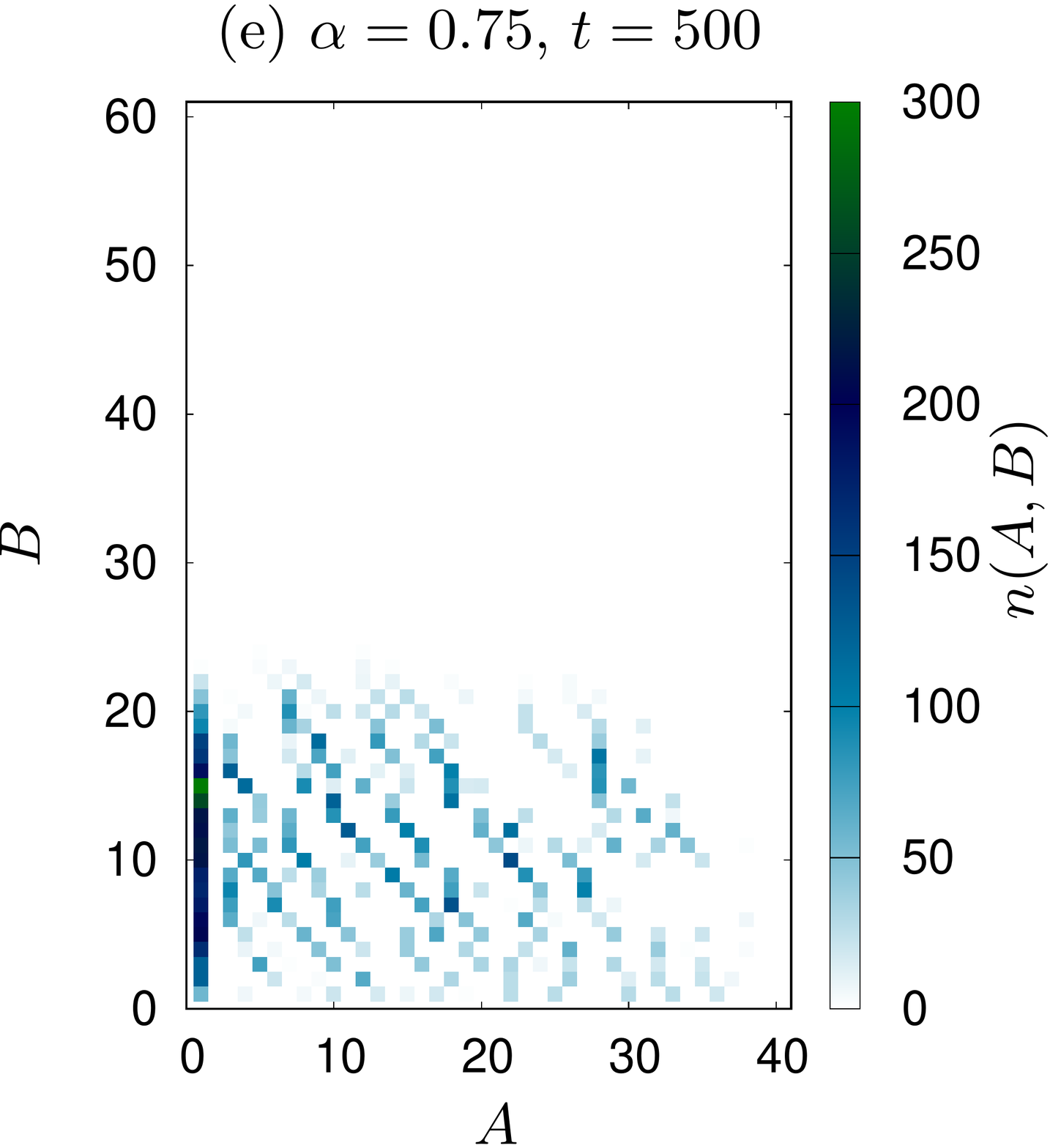}\hfill\includegraphics[width=0.30\textwidth]{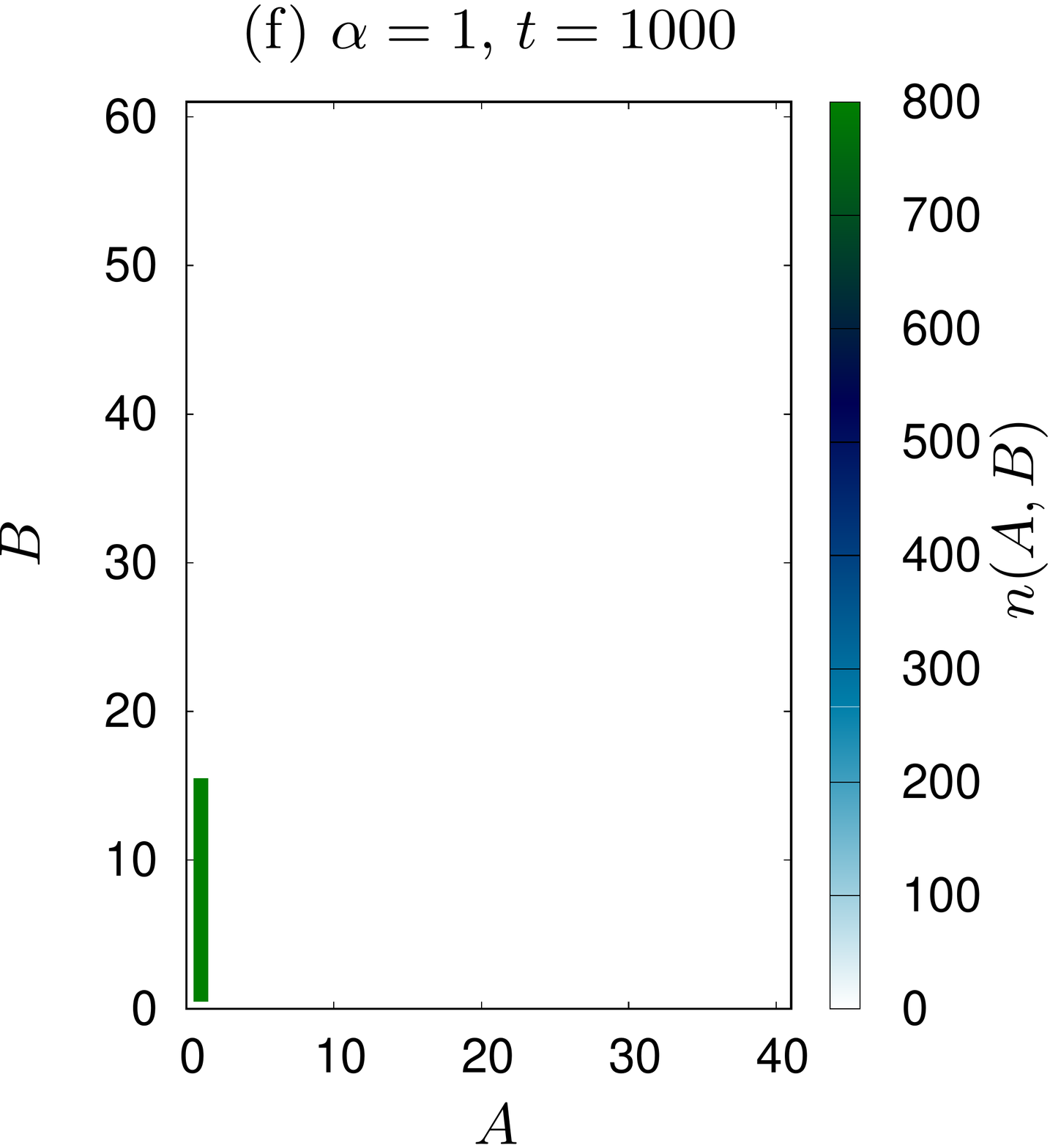}
\caption{\label{fig:synchronous}The distribution of the final number $n(A,B)$ of actors with status $(A,B)$ after $t$ time steps and various values of $\alpha$, $L_A=40$, $L_B=60$ and {\em synchronous} sites update.}
\end{figure*}

Two processes are competing in the time evolution. First is the fear-driven rejection, which takes place by a lowering of status $A$ (real status) of an actor by his direct neighbour with lower status $A$ and the same status $B$ (surface status). This process takes place with probability $\alpha$. The second process is an enhancement of status $B$ (surface status) of an actor by his direct neighbour with higher status $A$ and the same status $B$. This process takes place with probability $1-\alpha$. Actors with different status $B$ do not interact. The latter rule assures, that the action to increase the status $B$ of a neighbour neutralises his rejection and allows to preserve own status $A$. This rule is not possible if we have only one-dimensional status, as was assumed earlier \cite{m1,m2}.

Every site of rectangle lattice $\mathcal{G}=\{(A,B): 1\le A\le L_A, 1\le B\le L_B\}$ represents the number $n(A,B)$ of actors with real status $A$ and surface status $B$.
Initially, all sites for $B\le 15$ are occupied by twenty actors.
Every time step $t$, every pair of actors at position $\{ (A,B)\cup (A+1,B) \}$ for which $n(A,B;t)n(A+1,B;t)>0$ apply either SDS (with probability of $1-\alpha$):
\begin{subequations}
  \label{eq:sds}
  \begin{equation} n(A,B+1;t+1)=n(A,B+1;t)+1, \end{equation}
  \begin{equation} n(A,B;t+1)=n(A,B;t)-1 \end{equation}
\end{subequations}
or the fear-driven rejection process (with probability $\alpha$):
\begin{subequations}
  \label{eq:nosds}
  \begin{equation} n(A+1,B;t+1)=n(A+1,B;t)-1, \end{equation}
  \begin{equation} n(A,B;t+1)=n(A,B;t)+1. \end{equation}
\end{subequations}

In the asynchronous version, 
one time step is equivalent to an update of all pairs in random order. In both versions, the random number (to apply SDS or not) is selected for each pair separately.

\section{Results}

In the upper part of Tab.~\ref{tab:evolution} the links to videos showing status-time system evolution obtained with parallel update scheme are provided.

The final states of system evolution for various values of probabilities $\alpha$ and {\em synchronous} sites update are shown in Figs.~\ref{fig:synchronous}(b)--(f).
In Fig.~\ref{fig:synchronous}(a) the common initial state of the system is presented.
For $\alpha=0$ (Fig.~\ref{fig:synchronous}(b)) all actors apply SDS, which yields generation of plenty actors with high surface status $B$.
On contrary, assuming $\alpha=1$ pushes system to the final state presented in Fig.~\ref{fig:synchronous}(f) with all actors with minimal real status $A=1$ and the surface status $B$ the same as initial.
We note that the system evolution for $\alpha=0$ and $\alpha=1$ presented in Fig.~\ref{fig:synchronous}(b) and \ref{fig:synchronous}(f) are common for both schemes discussed in this paper.

\begin{figure*}
\includegraphics[width=0.32\textwidth]{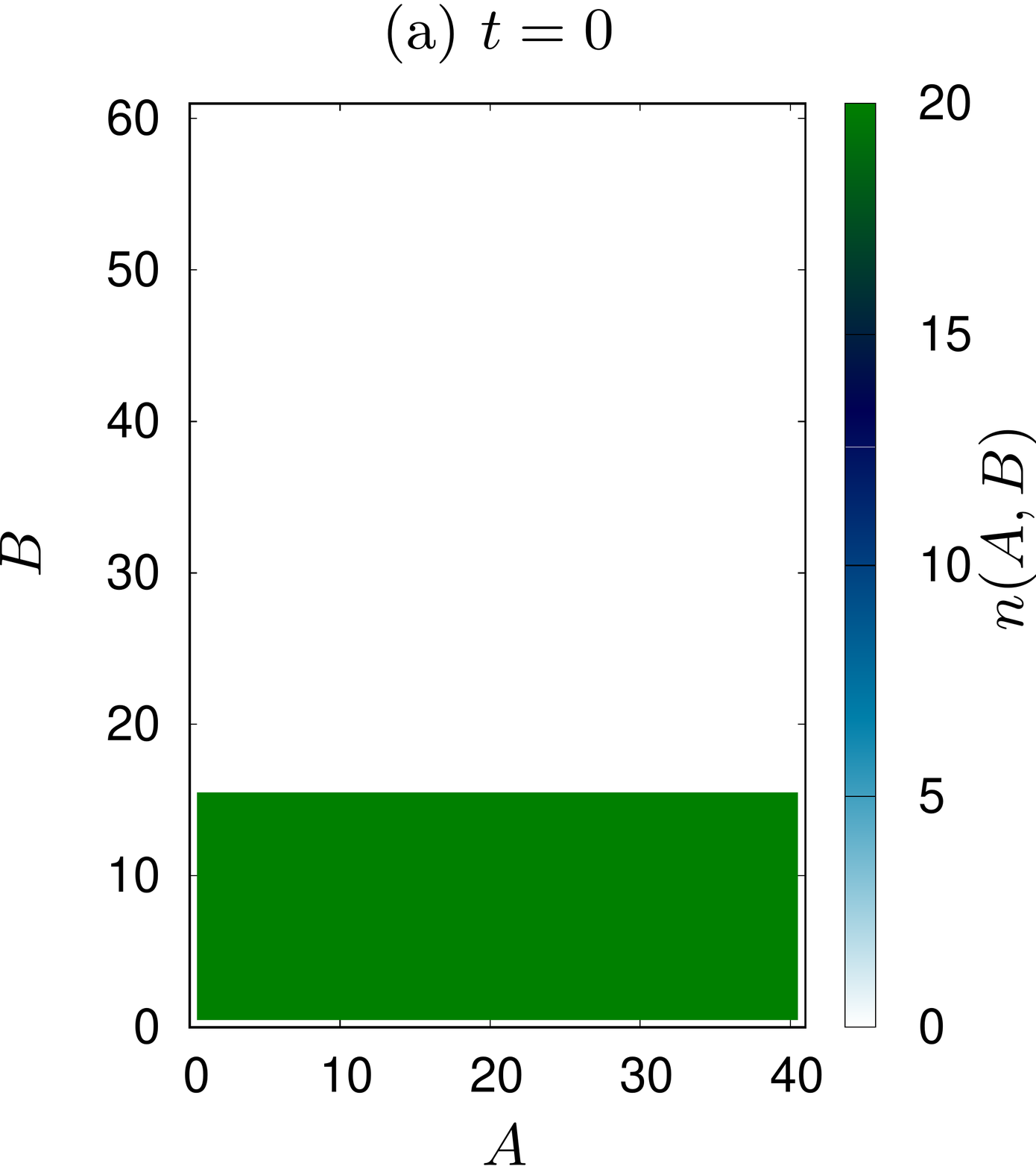}\hfill\includegraphics[width=0.32\textwidth]{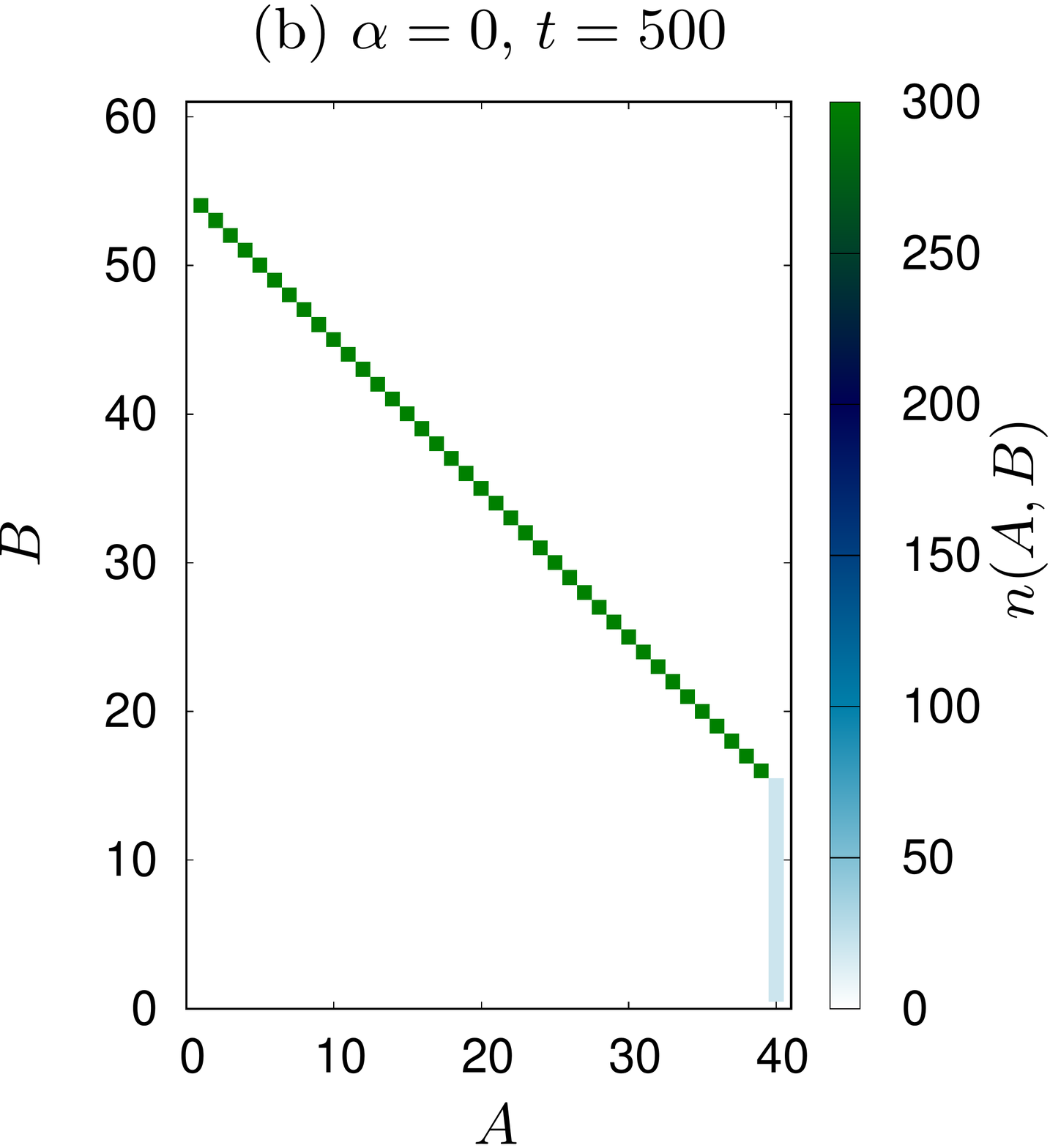}\hfill\includegraphics[width=0.32\textwidth]{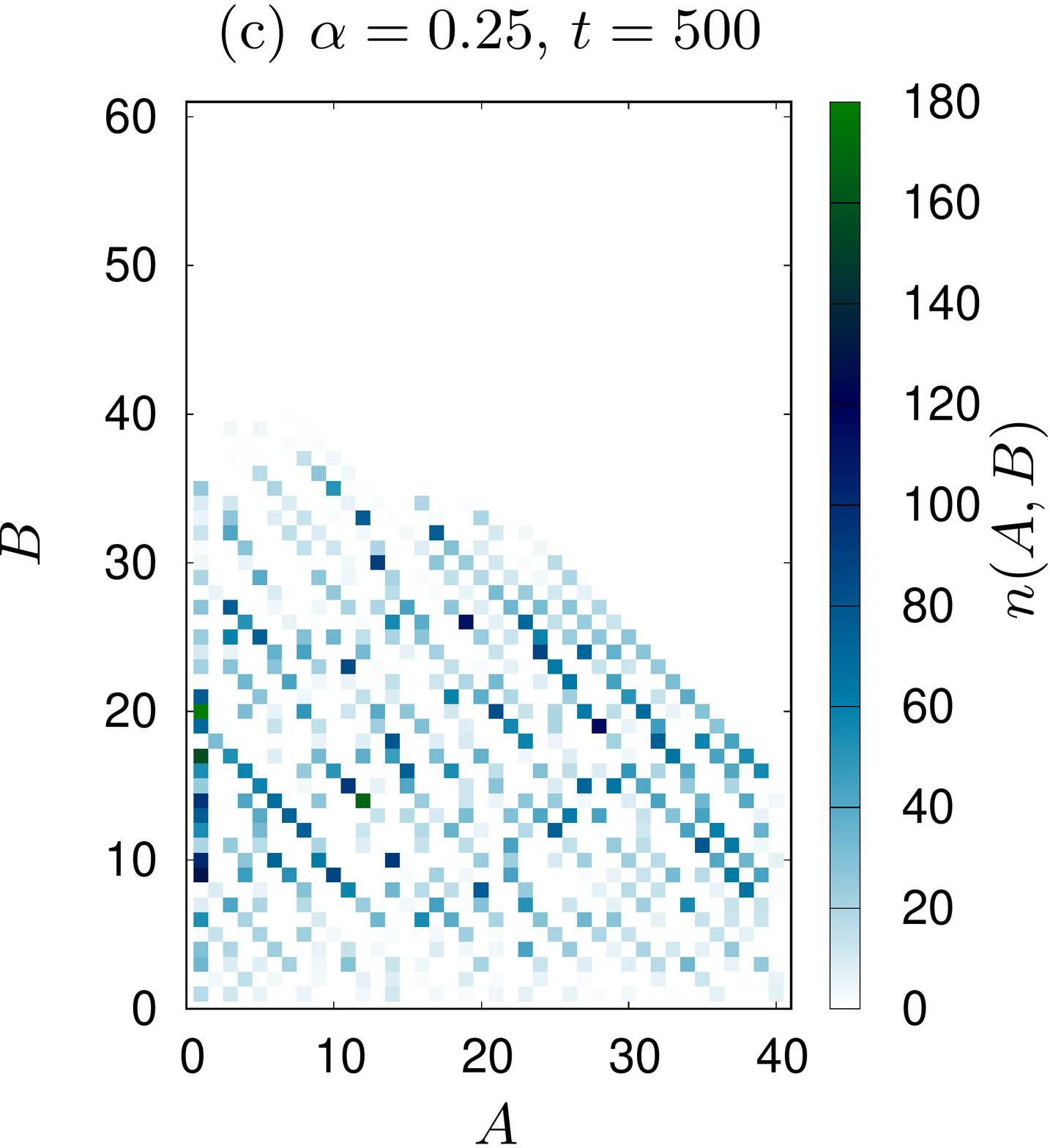}\\
\includegraphics[width=0.32\textwidth]{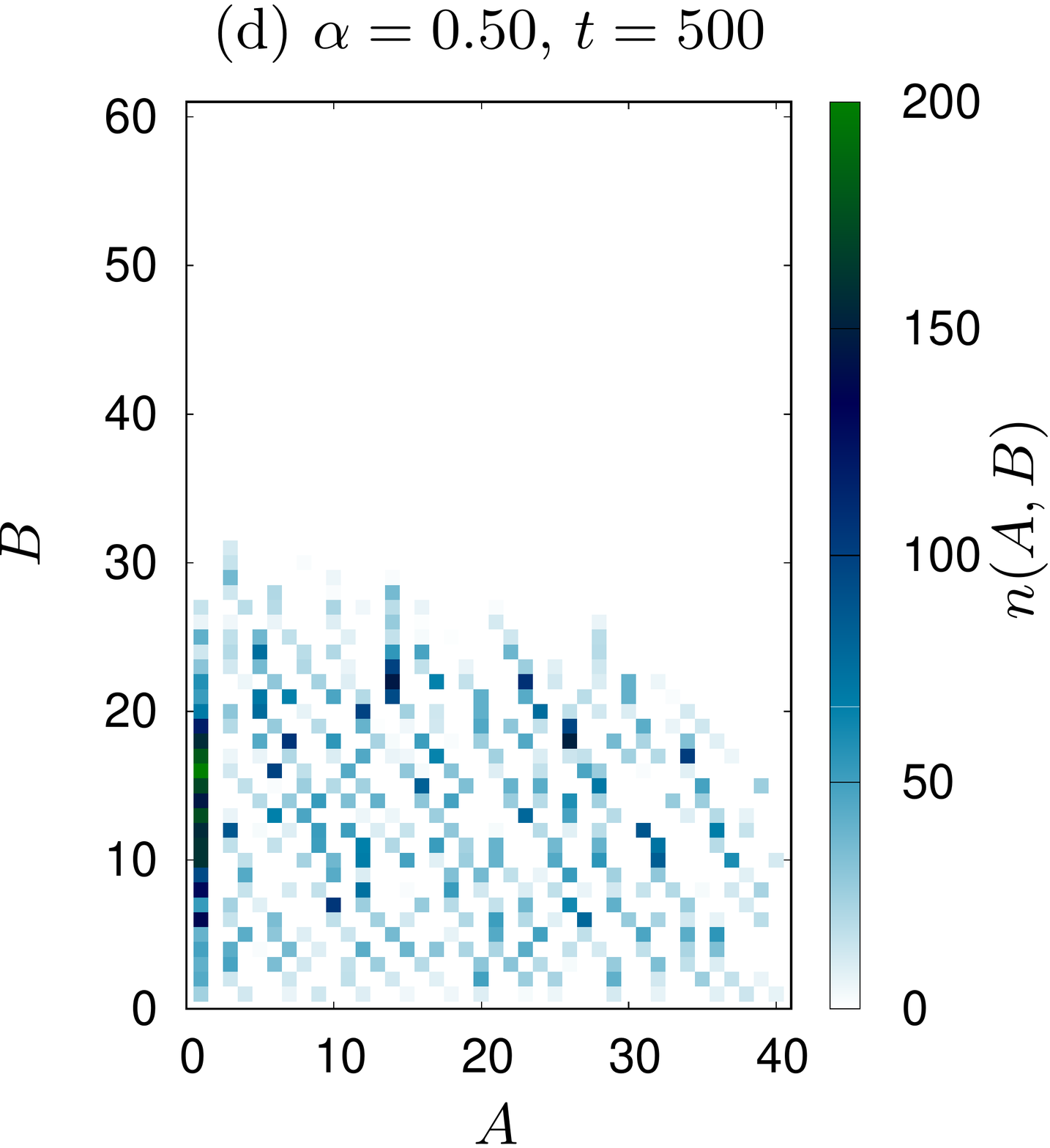}\hfill\includegraphics[width=0.32\textwidth]{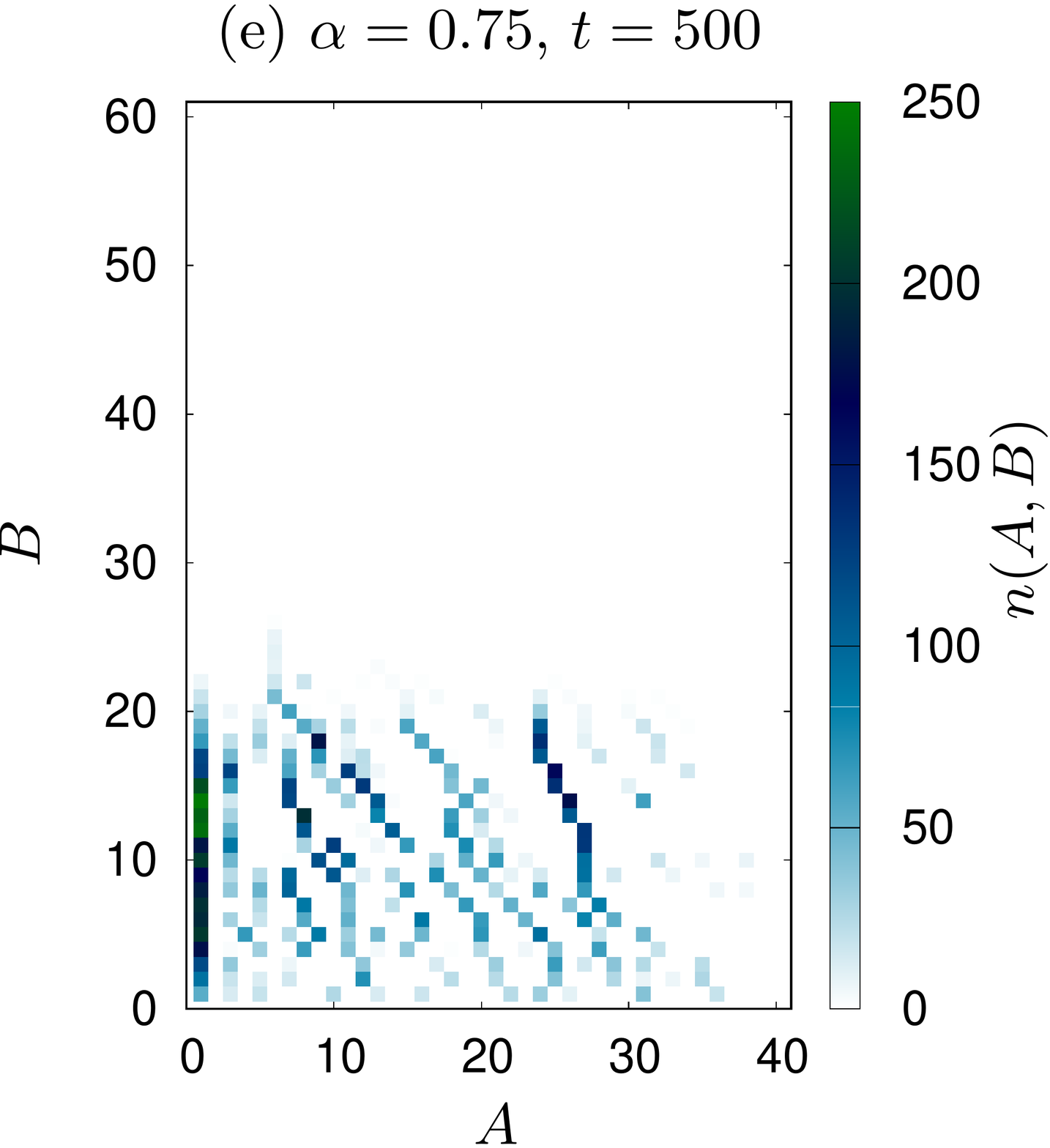}\hfill\includegraphics[width=0.32\textwidth]{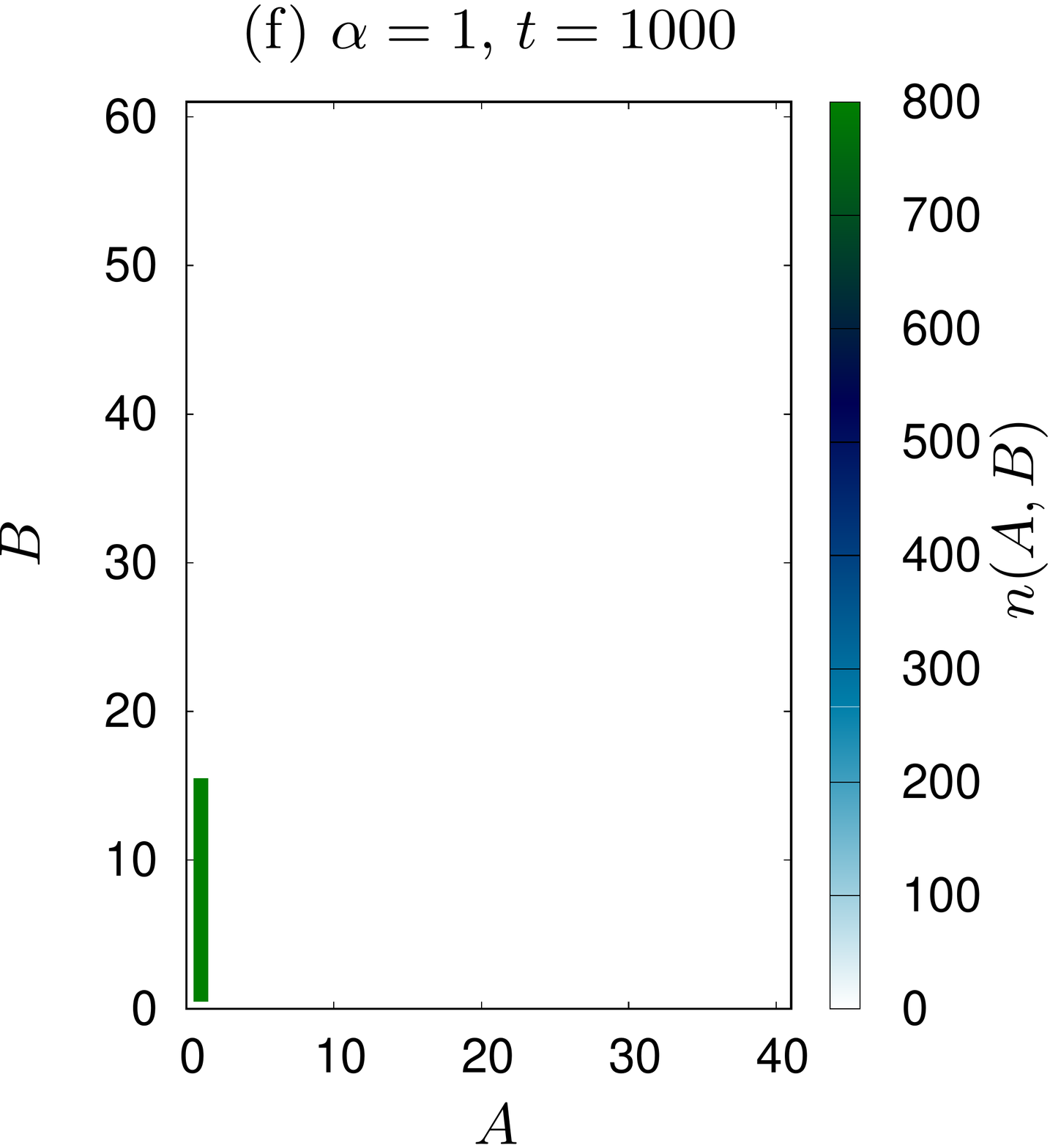}
	\caption{\label{fig:asynchronous}The distribution of the final number $n(A,B)$ of actors with status $(A,B)$ after $t$ time steps and various values of $\alpha$, $L_A=40$, $L_B=60$ and {\em asynchronous} sites update.}
\end{figure*}

In Figs.~\ref{fig:synchronous}(b)--(f) we see that the final outcome depends on the probability $\alpha$; indeed, the asymptotic (stationary) state depends only on one parameter $\alpha /(1-\alpha)$. For $\alpha$ close to one, SDS is not active, and the rejection reduces the real status A of all actors to the minimal value. In the opposite limit $\alpha$ close to zero, a straight line appears in the plane $(A,B)$, with a slope $\pi/4$ to both axes. This result is the same as the one obtained in Ref.~\onlinecite{1903.04291}. As we argued there, the angle has no real mining, because it is the consequence of the assumed scales of both statuses, both arbitrary and unverifiable.


In Figs.~\ref{fig:asynchronous}(b)--(f) the results of simulations for the {\em asynchronous} version of sites update are presented. As we see, the results for both schemes are the same up to details due to the randomness of the system. In both cases we observe characteristic chess-board-like pattern of neighboring cells. These configurations remain unchanged during the evolution; this is due to the rule that cells with different status 
$B$ do not interact anymore.

\section{Discussion}

The results on the number $n(A,B)$ of actors of status $(A,B)$ obtained within the synchronous (parallel) and asynchronous schemes are qualitatively the same as those produced by the differential equations \cite{1903.04291}. Namely, an enhancement of the surface status $B$ compensates a deficiency of the real one $A$. It appears that the localness is not crucial. This result is consistent with our previous tests \cite{1903.04291}, where the interaction strength was assumed to decrease with the difference of statuses. Both the present calculations for local interactions and the tests mentioned above take into account the hint by Blau \cite{Blau}, that the interaction strength decreases with the distance in the status space. Yet, the final results remain not influenced by the localness of the automaton rules.

As we mentioned in the Introduction, due to an asymmetry of automaton rules, the flow of actors is directional: actors move only downwards along $A$ axis and upwards along $B$ axis. This assumption leads to vanishing the fittest part of the population, i.e. those with the highest real status $A=L_x$. This phenomenon is described in the literature of genetics as an error catastrophe.  The error catastrophe is the first step in the so-called Muller's ratchet \cite{Muller-1964}
observed also in the Eigen's quasispecies \cite{Eigen-1989,*Tiggemann1998}. Next, also the secondary most fitted (i.e. with real status $A=L_x-1$) may disappear, what is the second step in the Muller's ratchet, etc. Due to the asymmetry of automaton rules the fraction with given $A$, once vanished, will never appear again.  As we can see in Fig.~\ref{fig:synchronous}(b) only taking by actors the pure SDS strategy (for $\alpha=0$) may prevent the error catastrophe. On the other hand, for $\alpha=1$ all but the last available step of Muller's ratchet take place when SDS is avoided. On the contrary to genetics, here the effect is due to a conscious action of actors with low status A. Notwithstanding, for $\alpha>0$ the catastrophe is unavoidable (see Figs.~\ref{fig:synchronous}(c)--(f), \ref{fig:asynchronous}(c)--(f)). 

Concluding, the outcome of the simulations indicates that a quick application of SDS by a smart actor blocks the fear-driven rejection. We can expect that in a consequence, SDS is simultaneously putting the subject actor in dependence of surface praising.

\bibliography{Blau,ca,km}{}


\end{document}